\documentstyle[12pt]{article}

\begin{document}
{\sf \begin{center} \noindent
{\Large \bf Plasma planar filaments instability and Alfven waves}\\[3mm]

by \\[0.3cm]

{\sl L.C. Garcia de Andrade}\\

\vspace{0.5cm} Departamento de F\'{\i}sica
Te\'orica -- IF -- Universidade do Estado do Rio de Janeiro-UERJ\\[-3mm]
Rua S\~ao Francisco Xavier, 524\\[-3mm]
Cep 20550-003, Maracan\~a, Rio de Janeiro, RJ, Brasil\\[-3mm]
Electronic mail address: garcia@dft.if.uerj.br\\[-3mm]
\vspace{2cm} {\bf Abstract}
\end{center}
\paragraph*{}
Inhomogeneous plasmas filaments instabilities are investigated by
using the techniques of classical differential geometry of curves
where Frenet torsion and curvature describe completely the motion of
curves. In our case the Frenet frame changes in time and also
depends upon the other coordinates taking into account the
inhomogeneity of the plasma. The exponential perturbation method so
commonly used to describe cosmological perturbatons is applied to
magnetohydrodynamic (MHD) plasma equations to find longitudinal
modes describing Alfven waves propagation modes describing plasma
waves in the medium. Stability is investigated in the imaginary axis
of the spectra of complex frequencies ${\omega}$ or
$Im(\omega)\ne{0}$.\vspace{0.5cm} \noindent {\bf PACS numbers:}
\hfill\parbox[t]{13.5cm}{02.40.Hw-classical differential geometry}

\newpage
\section{Introduction}
 The topology and geometry of hydrodynamical and MHD instability have been called \cite{1} one of the most important parts
 of plasma science. Arnold and Khesin \cite{2} have investigated the
 role of topology and Riemannian geometry to investigate MHD dynamos so important
 for use in geophysics and solar physics \cite{3}. Use of chaotic flows have been developed recently by Thiffeault and Boozer\cite{4}.
 Twisted filamentary magnetic structures have been applied in solar physics \cite{5} and in plasma
 filaments electric carrying-current loops \cite{5}. In this
 paper we consider the generalized filamentary structures and its
 unstable profiles. One of the simplest methods to investigate
 instabilities is the so-called exponential instability which is
 carachterized by the relation $Im(\omega)>0$ where Im denotes the
 imaginary part of complex structure of the spectra of
 perturbations where any physical quantity in equilibrium $Q_{0}$ is
 perturbed by a quantity
 $Q_{1}={Q_{1}}^{0}exp[-i({\omega}t-(k_{||}s+k_{\perp}n)]$ where s
 is the coordinate along the filament and n is along the filament
 direction. The quantities $k_{||}$ and $k_{\perp}$ represent the
 respective wave numbers of propagation. Thus Q could represent
 any perturbed physical quantity such as magnetic fields or flow
 speed. Throughout the paper we use the notation of a previously paper on vortex filaments in MHD \cite{6}. The paper is
 organized as follows: In section 2 we decompose MHD equations on a Frenet frame
 along the thin filament and perturb the MHD vector equations. In section 3 we compute the perturbations of the magnetic
 filament and analyse the stable and unstable modes and the transition to stable to unstable ones. In section 4 we present
 the conclusions.
 \section{Scalar perturbations in MHD filamentary structures}
 Let us now start by considering the MHD field equations
\begin{equation}
{\nabla}.\vec{B}=0 \label{1}
\end{equation}
\begin{equation}
{\nabla}{\times}{\vec{B}}= {\partial}_{t}\vec{B} \label{2}
\end{equation}
\begin{equation}
{\nabla}.({{\rho}\vec{v}})
+{\partial}_{t}{\rho} =0 \label{3}
\end{equation}
\begin{equation}
{\nabla}{\times}{(\vec{v}{\times}{\vec{B}})}
= {\partial}_{t}\vec{B}
\label{4}
\end{equation}
\begin{equation}
\frac{d}{dt}[{\frac{p}{{\rho}^{\gamma}}}]=0 \label{5}
\end{equation}
\begin{equation}
{\rho}\frac{d\vec{v}}{dt}=\vec{J}{\times}{\vec{B}}-{\nabla}p
\label{6}
\end{equation}
where the equilibrium quantities are
\begin{equation}
\vec{B}=B_{0}\vec{t} \label{7}
\end{equation}
\begin{equation}
{\vec{J}}_{0}= 0 \label{8}
\end{equation}
\begin{equation}
\vec{v_{0}}=0 \label{9}
\end{equation}
\begin{equation}
p=p_{0}+p_{1} \label{10}
\end{equation}
\begin{equation}
{\rho}={\rho}_{0}+{\rho}_{1}\label{11}
\end{equation}
magnetic field $\vec{B}$ along the filament is defined by the
expression $\vec{B}=B_{s,n,t}\vec{t}$ and $B_{s,t}$ is the component
along the arc length s of the filament depending upon time. Here we
consider that $B_{0}$ does not depend on time and also does not
depend on normal coordinates n, so $B_{0}(s)$. The vectors $\vec{t}$
and $\vec{n}$ along with binormal vector $\vec{b}$ together form the
Frenet frame which obeys the Frenet-Serret equations
\begin{equation}
\vec{t}'=\kappa\vec{n} \label{12}
\end{equation}
\begin{equation}
\vec{n}'=-\kappa\vec{t}+ {\tau}\vec{b} \label{13}
\end{equation}
\begin{equation}
\vec{b}'=-{\tau}\vec{n} \label{14}
\end{equation}
the dash represents the ordinary derivation with respect to
coordinate s, and $\kappa(s,t)$ is the curvature of the curve where
$\kappa=R^{-1}$. Here ${\tau}$ represents the Frenet torsion. We
follow the assumption that the Frenet frame may depend on other
degrees of freedom such as that the gradient operator becomes
\begin{equation}
{\nabla}=\vec{t}\frac{\partial}{{\partial}s}+\vec{n}\frac{\partial}{{\partial}n}+\vec{b}\frac{\partial}{{\partial}b}
\label{15}
\end{equation}
 The other equations for the other legs of the Frenet frame are
\begin{equation}
\frac{\partial}{{\partial}n}\vec{t}={\theta}_{ns}\vec{n}+[{\Omega}_{b}+{\tau}]\vec{b}
\label{16}
\end{equation}
\begin{equation}
\frac{\partial}{{\partial}n}\vec{n}=-{\theta}_{ns}\vec{t}-
(div\vec{b})\vec{b} \label{17}
\end{equation}
\begin{equation}
\frac{\partial}{{\partial}n}\vec{b}=
-[{\Omega}_{b}+{\tau}]\vec{t}-(div{\vec{b}})\vec{n}\label{18}
\end{equation}
\begin{equation}
\frac{\partial}{{\partial}b}\vec{t}={\theta}_{bs}\vec{b}-[{\Omega}_{n}+{\tau}]\vec{n}
\label{19}
\end{equation}
\begin{equation}
\frac{\partial}{{\partial}b}\vec{n}=[{\Omega}_{n}+{\tau}]\vec{t}-
\kappa+(div\vec{n})\vec{b} \label{20}
\end{equation}
\begin{equation}
\frac{\partial}{{\partial}b}\vec{b}=
-{\theta}_{bs}\vec{t}-[\kappa+(div{\vec{n}})]\vec{n}\label{21}
\end{equation}
The equations \cite{9} for the time evolution of the Frenet frame
yields
\begin{equation}
\dot{\vec{t}}= -{\tau}{\kappa}\vec{n}+ {\kappa}'\vec{b} \label{22}
\end{equation}
\begin{equation}
\dot{\vec{n}}=-{\kappa}{\tau}\vec{t} \label{23}
\end{equation}
\begin{equation}
\dot{\vec{b}}=-{\kappa}'\vec{t} \label{24}
\end{equation}
where ${\kappa}'=\frac{\partial}{{\partial}s}{\kappa}$.
\section{Unstable solutions of MHD plasma filaments and Alfven waves} Substitution of
the above equations into the LHS of magnetic equation reads
\begin{equation}
{\nabla}{\times}{\vec{B}_{0}}={\mu}_{0}{\vec{J}_{0}}=0\label{25}
\end{equation}
Expansion of this equation on the Frenet frame yields
\begin{equation}
{B}_{0}[\vec{n}{\times}{\partial}_{n}\vec{t}+\vec{b}{\times}{\partial}_{b}\vec{t}]=0\label{26}
\end{equation}
where we have used that $B_{0}$ is constant. Substitution of the
above dynamical relations for the Frenet frame yields the following
geometrical constraints
\begin{equation}
{\Omega}_{n}=-{\tau}=0\label{26}
\end{equation}
for planar filaments condition implies that torsion ${\tau}=0$. The
remaining constraint is
\begin{equation}
{\kappa}={\Omega}_{b} \label{27}
\end{equation}
where the ${\Omega}_{a}$ , where $(a=s,n,b)$ represent the
abnormalities \cite{10}. In particular if ${\Omega}_{s}=0$ we say
that the filament bundles are geodesic. Note also that we consider
that the curvature and torsion are perturbed but we assume that the
torsion remains zero after perturbation so the motion is constrained
to be perturbed in the plane. In mathematical terms,
${\tau}={\tau}_{0}+{\tau}_{1}$ and
${\kappa}={\kappa}_{0}+{\kappa}_{1}$ where ${\tau}_{0}=0$ and
${\tau}_{1}=0$. Since the current density is written as
$\vec{J}={\rho}{\vec{v}}$ one obtain
\begin{equation}
{\nabla}{\times}{\vec{B}_{1}}={\mu}_{0}{\vec{J}_{1}}\label{28}
\end{equation}
which yields
\begin{equation}
{\partial}_{s}{B}_{1}+[{\theta}_{ns}+{\theta}_{nb}]B_{1}=0\label{29}
\end{equation}
which reduces to
\begin{equation}
[ik_{\perp}-{\kappa}_{0}]B_{1}= {\mu}_{0}J_{1}\label{30}
\end{equation}
By calling ${\theta}:=[{\omega}t-(k_{||}s+k_{\perp}n)]$ , where here
${\omega}=Re{\omega}+iIm{\omega}$ and using the Moivre law
$exp[-i{\theta}]=cos{\theta}-isin{\theta}$ into equation (\ref{30})
yields
\begin{equation}
ik_{\perp}{B^{0}}_{1}[cos{\theta}-isin{\theta}]=({\kappa}_{0}{B^{0}}_{1}+{\mu}_{0}{J^{0}}_{1})[cos{\theta}-isin{\theta}]\label{31}
\end{equation}
This complex equation yields two scalar real equations which
solution is
\begin{equation}
[k_{\perp}{B^{0}}_{1}]^{2}=[{\kappa}_{0}{B^{0}}_{1}+{\mu}_{0}{J^{0}}_{1})]^{2}\label{32}
\end{equation}
Since $B_{0}(s)$ the other Maxwell equation ${\nabla}.\vec{B}=0$
becomes
\begin{equation}
{\partial}_{s}{B}_{0}+[{\theta}_{bs}+div\vec{b}]{B}_{0}=0 \label{33}
\end{equation}
which yields the solution
\begin{equation}
{B}_{0}=-c_{0}exp[\int{({\theta}_{bs}+{\theta}_{ns})ds}] \label{34}
\end{equation}
where $c_{0}$ is an integration constant. Now the perturbed equation
is
\begin{equation}
{\nabla}.\vec{{B}_{1}}=0 \label{35}
\end{equation}
which reduces to the expression
\begin{equation}
{\partial}_{s}{B}_{1}+[{\theta}_{bs}+div\vec{b}]{B}_{1}=0 \label{36}
\end{equation}
which in turn produces the following complex equation
\begin{equation}
ik_{||}{B^{0}}_{1}[cos{\theta}-isin{\theta}]=-{B^{0}}_{1}[cos{\theta}-isin{\theta}]\label{37}
\end{equation}
which being analogous to equation (\ref{31}) can be solved in the
same way to yield
\begin{equation}
k_{||}=\pm[{\theta}_{ns}+{\theta}_{sb}]\label{38}
\end{equation}
Now let us solve the conservation of mass solution as
\begin{equation}
i{\omega}{\rho}_{1}=v_{1}{\rho}_{0}div{\vec{b}}\label{39}
\end{equation}
Expanding this complex relation we are able to find out
\begin{equation}
-i[Re{\omega}+iIm{\omega}]{\rho}_{1}=v_{1}{\rho}_{0}div{\vec{b}}\label{40}
\end{equation}
which yields two real equations which together yields
\begin{equation}
[Re{\omega}{{\rho}^{0}}_{1}]^{2}=[Im{\omega}{{\rho}^{0}}_{1}+{v^{0}}_{1}{\rho}_{0}]^{2}\label{41}
\end{equation}
In the branch $Re{\omega}=0$ a simple solution of this equation
allows us to investigate the instability of the plasma filaments,
which is
\begin{equation}
Im{\omega}=\frac{{v^{0}}_{1}{\rho}_{0}}{{{\rho}^{0}}_{1}}\label{42}
\end{equation}
Note that in this branch exponential instability is possible since
$Im{\omega}>0$ implies
\begin{equation}
\frac{{v^{0}}_{1}{\rho}_{0}}{{{\rho}^{0}}_{1}}>0\label{43}
\end{equation}
Since the mass densities of the fluid are always positive the
instabilities imposes constraints on the velocity ${v^{0}}_{1}>0$,
thus if the velocity is negative or attractive the plasma filament
is stable and the amplification of the magnetic field as happens in
dynamos is not possible and magnetic field is damped. Let us now to
investigate the remaining Maxwell MHD equations. They are
\begin{equation}
{\omega}_{0}:=Im{\omega}=\pm{\frac{k_{||}L{B_{0}}({\kappa}_{0}+div{\vec{n}})}
{{B^{0}}_{1}}}
 \label{44}
\end{equation}
where $\int{ds}=L$ which is the length of the filament. In the case
of solar loops for example $L={\pi}R$ by considering that the half
of the solar filament is under the surface of the Sun. Writhing the
expression for the Alfven wave frequency as
${{\omega}_{0}}^{2}=[k_{||}V_{a}]^{2}$ and comparing it with the
expression (\ref{44}) one obtains the Alfven velocity as
\begin{equation}
{V_{a}}^{2}=[\frac{LB_{0}({\kappa}_{0}+div{\vec{n}})}{{B^{0}}_{1}}]^{2}
 \label{45}
\end{equation}
To simplify our physical analysis we assume that the plasma filament
bundle obeys the relation $div{\vec{n}}=0$ which reduces expression
(\ref{45}) to
\begin{equation}
{V_{a}}=\pm[\frac{LB_{0}{\kappa}_{0}}{{B^{0}}_{1}}]
 \label{46}
\end{equation}
once we have taken the plus sign or example we show that Alfven
waves propagates along the ilamet with the same sign of velocity as
the one of the Frenet curvature of the equilibrium. The very last
equation yields a relation etween components of the pressure as
\begin{equation}
\frac{{{\rho}^{0}}_{1}}{{\rho}_{0}}p_{0}={{p}^{0}}_{1}\label{47}
\end{equation}
The solution described here is also well suitable for plasma
filaments tokamaks where the curvature is not perturbed since is
fixed by the topology of the plasma device.
\section{Conclusions}
 In conclusion, plasma MHD instability is investigated in the framework of the Frenet inhomogeneous frame. Alfven waves are
 found, where the velocity is expressed in terms of the Frenet curvature of planar filaments. This effect plays an important
 role in the construction of tokamaks and other plasma devices. Amplification of the magnetic fields is possible in the case of
 unstable filaments. Future work in the field of perturbations would include the plasma metric perturbations.

 \section*{Acknowledgements}
 Thanks are due to CNPq and UERJ for financial supports.

\end{document}